\documentclass{ws-procs9x6}            
\newcommand{\be}{\begin{equation}}
\newcommand{\ee}{\end{equation}}
\newcommand{\ba}{\begin{array}}
\newcommand{\ea}{\end{array}}
\newcommand{\bqa}{\begin{eqnarray}}
\newcommand{\eqa}{\end{eqnarray}}

\begin{document}
\title{Understanding $X(3872)$ and its decays in the extended Friedrichs scheme}

\author{Meng-Ting Yu and Zhi-Yong Zhou$^*$}

\address{School of Physics, Southeast University, Nanjing 211189,
People's Republic of China\\
$^*$E-mail: zhouzhy@seu.edu.cn}

\author{Zhiguang Xiao}

\address{Interdisciplinary Center for Theoretical Study, University of Science
and Technology of China, Hefei, Anhui 230026, China\\
E-mail: xiaozg@ustc.edu.cn}

\begin{abstract}
We present that the $X(3872)$ could be represented as a dynamically generated state in the extended Friedrichs scheme, in which the
ratio of ``elementariness" and  ``compositeness" of the different components in the $X(3872)$ is about
$Z_{c\bar c}:X_{\bar D^0 D^{0*}}: X_{ D^+ D^{-*}}: X_{\bar D^* D^*}$ $= 1:(2.67\sim 8.85):(0.45\sim 0.46):0.04$. Furthermore, its decays to  $\pi^0$ and a $P$-wave
charmonium $\chi_{cJ}$ state with $J=0,1$, or $2$, $J/\psi\pi^+\pi^-$, and $J/\psi\pi^+\pi^-\pi^0$ could be calculated out with the help of Barnes-Swanson model. The isospin breaking effects is easily understood in this scheme.
This calculation also shows that the decay rate of $X(3872)$
to $\chi_{c1}\pi^0$ is much smaller than its decay
rate to $J/\psi\pi^+\pi^-$.
\end{abstract}

\keywords{Friedrichs model; exotic state; charmonium; molecular state.}

\bodymatter

\section{Introduction}
The $X(3872)$ state, first discovered by Belle~\cite{Choi:2003ue},
has several enigmatic properties including its mass, narrow width,
isospin breaking effects in $X(3872)\rightarrow J/\psi\pi^+\pi^-$ and
$J/\psi\pi^+\pi^-\pi^0$ ~\cite{delAmoSanchez:2010jr}, as discussed in
the literature.  These properties of $X(3872)$ might be explained by
describing it as a dynamically generated state due to the coupling of
the bare $\chi_{c1}(2P)$ and $D\bar D^*$, $D^*\bar D^*$  continuum
states in the extended Friedrichs
scheme~\cite{Xiao:2016wbs,Zhou:2017dwj}.  Recently,
BESIII  searched
for the $X(3872)$ signals in $e^+e^-\to\gamma\chi_{cJ}\pi^0$
($J=0,1,2$) and reported an observation of $X(3872)\to\chi_{c1}\pi^0$
with a ratio of branching fractions~\cite{Ablikim:2019soz}
$\frac{\mathcal{B}(X(3872)\to\chi_{c1}\pi^0)}{\mathcal{B}(X(3872)\to
J/\psi\pi^+\pi^-)}=0.88^{+0.33}_{-0.27}\pm 0.10$.  Soon after, Belle
made a search for $X(3872)$ in $B^+\to \chi_{c1}\pi^0K^+$ but only
reported an upper limit\cite{Bhardwaj:2019}
$\frac{\mathcal{B}(X(3872)\to\chi_{c1}\pi^0)}{\mathcal{B}(X(3872)\to
J/\psi\pi^+\pi^-)}<0.97$ at $90\%$ confidence level.  We will present
that by using the wave function of $X(3872)$ from the extended
Friedrichs scheme and the Barnes-Swanson constituent interchange
model~\cite{Barnes:1991em}, the decays of $X(3872)$ to different
hadronic final states could also be obtained.
\section{ Wave function of $X(3872)$ in the extended Friedrichs scheme}\label{efs}

In 1948, Friedrichs proposed an exactly solvable model to understand
an unstable state~\cite{Friedrichs:1948}, in which the free Hamiltonian $H_0$ has a bare continuous spectrum
$[E_{th},\infty)$, and a discrete eigenvalue $E_0$ imbedded in this
continuous spectrum~($E_0 >E_{th}$). The interaction Hamiltonian $V$
couples the bare continuous state and the discrete  state of
$H_0$ such that the discrete state is dissolved in the continuous state
and a resonance is produced.

This model was developed to a general form with many discrete and
continuous states~\cite{Xiao:2016mon}, and furthermore, some general
properties of the Friedrichs model, such as the wave functions of
resonance and the completeness relations, are
obtained~\cite{Xiao:2016dsx,Xiao:2016wbs}.  The free Hamiltonian $H_0$
can be expressed as $H=E_0 |0\rangle\langle 0| +\sum_{n,S,L}
\int_{E_{th,n}}^\infty \mathrm dE\, E |E;n,SL\rangle \langle E;n,SL|$ and the interaction Hamiltonian $V=\sum_{n,S,L}\int_{E_{th,n}}^\infty \mathrm{d}E  f^n_{SL}(E)|0\rangle\langle
E;n,SL|+h.c.,$
where $E_0$ denotes the bare mass of the discrete state,  $n$ the
``$n$" continuum state, $E_{th,n}$ the energy threshold of related
continuum state, $S$ and $L$ the total spin and the angular momentum
of the continuum states, $f^n_{SL}$ the coupling functions between the
bare state and the ``$n$" continuum state with particular $S,L$
quantum numbers. The eigenvalue problem of the full Hamiltonian $H=H_0+V$ is exactly solvable as mentioned above.
The eigenvalues of  bound states, virtual states or resonant
states could be found by solving $\eta(z)=0$
on the complex energy plane where
$\eta(z)=z-E_0-\sum_n\int_{E_{th,n}}^\infty\frac{\sum_{S,L}|f_{SL}^n(E)|^2}{z-E
}\mathrm{d}E$.
The form factor $f_{SL}$ is provided by some particular model, such as
the quark pair creation model~\cite{Micu:1968mk,Zhou:2017dwj}. By
using the results from  the famous Godfrey-Isgur model~\cite{Godfrey:1985xj} as the input, the
$X(3872)$ state is dynamically generated by the coupling between the
bare discrete $\chi_{c1}(2P)$ state and the continuum $D\bar D^*$ and $D^*\bar D^*$ states~\cite{Zhou:2017dwj}, and its wave function  could be explicitly written down as
\begin{align}
&|X\rangle=N_B\Big(|c\bar c\rangle+\int_{M_{00}}^\infty \mathrm{d}E
\sum_{l,s}\frac{f^{00}_{ls}(E)} {z_X-E}(|E\rangle_{ls}^{D^0\bar
D^{0*}}+C.C.
)\nonumber\\&+\int_{M_{+-}}^\infty
\mathrm{d}E\sum_{l,s}\frac{f^{+-}_{ls}(E)}
{z_X-E}(|E\rangle_{ls}^{D^+D^{-*}}+C.C.)+\cdots\Big) ,
\label{eq:X3872-wave-funtion}
\end{align}
where $C.C.$ means the
corresponding charge conjugate state, $|c\bar c\rangle$ denotes the bare $\chi_{c1}(2P)$ state and
$|E\rangle_{ls}^n=\sqrt{\mu k}|k, j\sigma,ls\rangle$ denotes  the
two-particle $``n"$ state with the reduced mass $\mu$, the
magnitude of one-particle three-momentum $k$ in their $c.m.$ frame,
total spin $s$, relative orbital angular momentum $l$, total angular
momentum $j$, and its third component $\sigma$.
Based on the probability explanation of the wave function, one could obtain the ``elementariness" and ``compositeness" of $X(3872)$ as
$Z_{c\bar c}:X_{\bar D^0 D^{0*}}: X_{ D^+ D^{-*}}: X_{\bar D^* D^*}$ $= 1:(2.67\sim 8.85):(0.45\sim 0.46):0.04$, which means $\bar D^0 D^{0*}$ component is dominant in the $X(3872)$ state.

\section{Decays of $X(3872)$}

In calculating the decays of the $X(3872)$ to $\chi_{cJ}\pi^0$ for $J=0, 1, 2$, the partial-wave decay amplitude reads
\begin{align}
&F_{l's'}={_{l's'}}\langle \chi_{cJ}\pi^0|H_I|X(3872)\rangle=N_B\Big({^{\chi_{cJ}\pi^0}_{\ ~ l's'}}\langle E'|H_I|c\bar c\rangle\nonumber\\
&+\int_{M_{00}}^\infty \mathrm{d}E
\sum_{l,s}\frac{f^{00}_{ls}(E)} {z_X-E}({^{\chi_{cJ}\pi^0}_{\ ~ l's'}}\langle E'|H_I|E\rangle_{ls}^{D^0\bar
D^{0*}}+C.C.
)\nonumber
\\ &+\int_{M_{+-}}^\infty
\mathrm{d}E\sum_{l,s}\frac{f^{+-}_{ls}(E)}
{z_X-E}({^{\chi_{cJ}\pi^0}_{\ ~ l's'}}\langle
E'|H_I|E\rangle_{ls}^{D^+D^{-*}}+C.C.)+\cdots\Big).
\label{eq:decay-amplitude}
\end{align}
Once the matrix elements for $D\bar D^*\to \chi_{cJ}\pi^0$  are obtained, the partial decay widths and
branching ratios are easily calculated directly.

The scattering amplitude of $D\bar D^*\to \chi_{cJ}\pi^0$ could be
calculated by using the Barnes-Swanson
model~\cite{Barnes:1991em,Barnes:2000hu}, in which the meson meson
scatterings are induced by the quark~(antiquark)-quark~(antiquark)
interactions. Such interactions are supposed to come from the one
gluon exchange~(OGE) interaction and the confinement interaction.
Four kinds of diagrams are considered, among which the quark-antiquark
interactions are denoted as $C_1$, $C_2$, and the
quark-quark(antiquark-antiquark) interactions are denoted as
$T_1$, and $T_2$. The other four ``post'' diagrams are also considered
similarly and averaged to reduce the so-called
``prior-post" ambiguity.

The partial wave scattering
amplitude for each diagram with only meson $C$ being a
$P$-wave state can be obtained by using
\begin{align}
&\mathcal{M}^1_{l'j_C,lj_B}=\sqrt{\mu k\mu' k'}\sum_{mm'm_{l_C}}
\langle j_B-m lm|10\rangle \langle j_C -m' l'm'|10\rangle\nonumber\\
& \times \langle l_C m_{l_C} s_C (-m'-m_{l_C})|j_C -m'\rangle \langle \phi_{14}\phi_{32}|\phi_{12}\phi_{34}\rangle\langle \omega_{14}\omega_{32}|H_C|\omega_{12}\omega_{34}\rangle\nonumber\\
&\times\int d\Omega_k\int d\Omega_{k'}\langle \chi_C\chi_D|I_{Space}^{m_{l_C}}[\vec k,\vec k']|\chi_A\chi_B\rangle Y_l^{m}(\hat k)Y_{l'}^{m'*}(\hat k')
\end{align}
where  $\langle \phi_{14}\phi_{32}|\phi_{12}\phi_{34}\rangle$ is the
flavor factor, and $\langle
\omega_{14}\omega_{32}|H_C|\omega_{12}\omega_{34}\rangle$ the color
factor, which is $-4/9$ and $4/9$ for interactions of $q\bar q$ and $q
q$ respectively. $\chi_A$ represents the spin wave function of meson
$A$. The spatial integral is
\begin{align}\label{spacefactor}
I_{Space}^{m_{l_C}}[\vec k,\vec k']=\int d^3p\int d^3q  \psi^A_{000}(\vec{p}_A)\psi^B_{000}(\vec{p}_B) \psi^{C*}_{01m_{l_C}}(\vec{p}_C)\psi^{D*}_{000}(\vec{p}_D)T_{fi}
\end{align}
where $\psi_{n_rLm_L}(\vec p_r)$ is the wave
function for the bare meson state, with $n_r$ being the radial quantum
number, $L$  the relative angular momentum of the quark and
anti-quark, $m_L$ its third component, $\vec p_r$ the relative momentum of quark and antiquark in the meson.
The quark interactions of spin-spin, color Coulomb, linear, OGE spin-orbit, linear spin-orbit, and OGE tensor terms are respectively expressed as
\bqa
T_{fi}=\left\{\begin{array}{c}
                -\frac{8\pi\alpha_s}{3m_1m_2}[\vec S_1\cdot\vec S_2] \\
                \frac{4\pi\alpha_s}{q^2}\mathbf{\textit{I}}  \\
                \frac{6\pi b}{q^4}\mathbf{\textit{I}}\\
                \frac{4i\pi\alpha_s}{q^2}\{\vec S_1\cdot[\vec q\times(\frac{\vec p_1}{2m_1^2}-\frac{\vec p_2}{m_1m_2})]+\vec S_2\cdot[\vec q\times(\frac{\vec p_1}{m_1m_2}-\frac{\vec p_2}{2m_2^2})]\}\\
                -\frac{3i\pi b}{q^4}[\frac{1}{m_1^2}\vec S_1\cdot(\vec q\times\vec p_1)-\frac{1}{m_2^2}\vec S_2\cdot(\vec q\times\vec p_2)]  \\
                \frac{4\pi\alpha_s}{m_1m_2q^2}[\vec S_1\cdot\vec q\vec S_2\cdot\vec q-\frac{1}{3}q^2\vec S_1\cdot\vec S_2]
              \end{array}\right.\nonumber\\
\eqa
where $\alpha_s=\sum_k \alpha_k e^{-\gamma_k  q^2}$ is the same
parametrization of the strong coupling as the
in the GI model and $m_1$ and $m_2$ are the masses of the two interacting quarks.

The decay amplitude of $X(3872)\to J/\psi\rho$ and $J/\psi\omega$  is much simpler because there is only $S$-wave states involved in the scattering amplitudes.

Finally, we find that the decay rates of $X(3872)$ to $\chi_{cJ}\pi^0$
for $J=0,1,2$ turn out to be one order of magnitude smaller than that to $J/\psi\pi^+\pi^-$.  Our result is
smaller than the central value measured by
BESIII~\cite{Ablikim:2019soz}, but we noticed that the result of
BESIII has sizable uncertainties, and more data are needed to increase
the statistics and reduce the error bar. The ratio  $\frac{\mathcal{B}(X(3872)\to
J/\psi\pi^+\pi^-\pi^0)}{\mathcal{B}(X(3872)\to J/\psi\pi^+\pi^-)}$ in
our calculation is about 1.6~\cite{Zhou:2019swr}, which is consistent with our previous calculation~\cite{Zhou:2017txt} and the experiment measurements~\cite{delAmoSanchez:2010jr}.
\section{Ackownledgement}
This work is supported by China National Natural
Science Foundation under contract  No. 11975075, No. 11575177, and No. 11105138, and  the Natural Science Foundation of
Jiangsu Province of China under contract No. BK20171349.


\bibliographystyle{ws-procs9x6} 
\bibliography{Ref}

\begin{thebibliography}{10}

\bibitem{Choi:2003ue}
S.~K. Choi {\em et~al.}, {Observation of a narrow charmonium - like state in
  exclusive $B^{\pm} \to K^{\pm} \pi^+ \pi^- J / \psi$ decays}, {\em Phys. Rev.
  Lett.} {\bf 91}, p. 262001  (2003).

\bibitem{delAmoSanchez:2010jr}
P.~del Amo~Sanchez {\em et~al.}, {Evidence for the decay X(3872)$\to J/\psi
  \omega$}, {\em Phys. Rev.} {\bf D82}, p. 011101  (2010).

\bibitem{Xiao:2016wbs}
Z.~Xiao and Z.-Y. Zhou, {On Friedrichs Model with Two Continuum States}, {\em
  J. Math. Phys.} {\bf 58}, p. 062110  (2017).

\bibitem{Zhou:2017dwj}
Z.-Y. Zhou and Z.~Xiao, {Understanding $X(3862)$, $X(3872)$, and $X(3930)$ in a
  Friedrichs-model-like scheme}, {\em Phys. Rev.} {\bf D96}, p. 054031  (2017),
  [Erratum: Phys. Rev. D 96, 099905 (2017)].

\bibitem{Ablikim:2019soz}
M.~Ablikim {\em et~al.}, {Observation of the decay $X(3872) \to \pi^0
  \chi_{c1}(1P)$}  (2019).

\bibitem{Bhardwaj:2019}
V.~Bhardwaj {\em et~al.}, {Search for X(3872) and X(3915) decay into
  $\chi_{c1}\pi^0$ in B decays at Belle}  (2019).

\bibitem{Barnes:1991em}
T.~Barnes and E.~S. Swanson, {A Diagrammatic approach to meson meson scattering
  in the nonrelativistic quark potential model}, {\em Phys. Rev.} {\bf D 46},
  131  (1992).

\bibitem{Friedrichs:1948}
K.~O. Friedrichs, On the perturbation of continuous spectra, {\em Commun. Pure
  Appl. Math.} {\bf 1}, 361  (1948).

\bibitem{Xiao:2016mon}
Z.~Xiao and Z.-Y. Zhou, {Partial Wave Decomposition in Friedrichs Model With
  Self-interacting Continua}, {\em J. Math. Phys.} {\bf 58}, p. 072102  (2017).

\bibitem{Xiao:2016dsx}
Z.~Xiao and Z.-Y. Zhou, {Virtual states and the generalized completeness
  relation in the Friedrichs model}, {\em Phys. Rev.} {\bf D 94}, p. 076006
  (2016).

\bibitem{Micu:1968mk}
L.~Micu, {Decay rates of meson resonances in a quark model}, {\em Nucl. Phys.}
  {\bf B10}, 521  (1969).

\bibitem{Godfrey:1985xj}
S.~Godfrey and N.~Isgur, {Mesons in a Relativized Quark Model with
  Chromodynamics}, {\em Phys. Rev.} {\bf D 32}, 189  (1985).

\bibitem{Barnes:2000hu}
T.~Barnes, N.~Black and E.~S. Swanson, {Meson meson scattering in the quark
  model: Spin dependence and exotic channels}, {\em Phys. Rev.} {\bf C63}, p.
  025204  (2001).

\bibitem{Zhou:2019swr}
Z.-Y. Zhou, M.-T. Yu and Z.~Xiao, {On decays of $X(3872)$ to $\chi_{cJ}\pi^0$
  and $J/\psi\pi^+\pi^-$}, {\em Phys. Rev.} {\bf D100}, p. 094025  (2019).

\bibitem{Zhou:2017txt}
Z.-Y. Zhou and Z.~Xiao, {Comprehending Isospin breaking effects of $X(3872)$ in
  a Friedrichs-model-like scheme}, {\em Phys. Rev.} {\bf D97}, p. 034011
  (2018).

\end{thebibliography}

\end{document}